\documentstyle[epsf,epsfig,wrapfig,here,12pt]{article}    
\begin{document}           
\baselineskip=0.33333in
\begin{quote} \raggedleft TAUP 2844-2006
\end{quote}
\vglue 0.5in
\begin{center}{\bf The Physical Meaning of Gauge Transformations \\
in Classical Electrodynamics}
\end{center}
\begin{center}E. Comay$^*$
\end{center}

\begin{center}
School of Physics and Astronomy \\
Raymond and Beverly Sackler Faculty of Exact Sciences \\
Tel Aviv University \\
Tel Aviv 69978 \\
Israel
\end{center}
\vglue 0.5in
\vglue 0.5in
\noindent
\vglue 0.2in
\noindent
Abstract:

The structure of electrodynamics based on the variational principle
together with causality and space-time homogeneity is analyzed. It
is proved that in this case the 4-potential is defined uniquely.
Therefore, the approach where Maxwell equations and the Lorentz
law of force are regarded as cornerstones of the theory is {\em not
equivalent} to the one described above.

\newpage

One may regard the equations of motion of a physical system as the
fundamental elements of a theory. Thus, the equations of motion can
be used for deriving useful formulas that describe properties of
the system. However, it is now acceptable that other principles
take a more profound role. Using this approach, the variational
principle, causality and homogeneity of space-time are regarded
here as the basis of the discussion. In particular, the fundamental
equations of motion of classical electrodynamics, namely, Maxwell
equations and the Lorentz law of force can be derived from the
variational principle [1,2]. The discussion carried out here proves
that in the case of electrodynamics, the two approaches are
{\em not equivalent} and that the variational principle imposes
further restrictions on the theory's structure.

It is proved in this work that if one adheres to the variational
principle together with causality and space-time homogeneity
then the 4-potential of electrodynamics
is defined uniquely. Therefore, in this approach, gauge
transformations are no more than useful mathematical tricks applied in
a process of solving specific problems.

The Lagrangian density used for a derivation of Maxwell equations is
(see [1], pp. 73-74; [2], pp. 596-597)
\begin{equation}
{\mathcal L} =
- \frac {1}{16\pi }F^{\mu \nu }F_{\mu \nu } - j^\mu A_\mu
\label{eq:LAGR}
\end{equation}

   In the present work, units where the speed of light
$c = 1$ and $\hbar = 1$ are used. Thus, one kind of dimension
exists and the length $[L]$ is used for this purpose.
Greek indices run from 0 to 3. The metric is diagonal and its entries are
(1,-1,-1,-1). The symbol $_{,\mu }$ denotes the partial differentiation
with respect to $x^\mu $. $A_\mu $ denotes the 4-potential and $F^{\mu \nu }$
denotes the antisymmetric tensor of the electromagnetic fields
\begin{equation}
F^{\mu \nu } = g^{\mu \alpha}g^{\nu \beta}(A_{\beta ,\alpha} -
A_{\alpha ,\beta})
=\left(
\begin{array}{cccc}
0   & -E_x & -E_y & -E_z \\

E_x &  0   & -B_z &  B_y \\

E_y &  B_z &   0  & -B_x \\

E_z & -B_y &  B_x &  0
\end{array}
\right).
\label{eq:FMUNU}
\end{equation}

For the simplicity of the discussion, let us examine the fields
associated with one charged particle $e$ whose motion is given.
This approach can be justified because, due to
the linearity of Maxwell equations, one finds that the fields of
a system of charges is a superposition of the fields of each
individual charge belonging to the system.
Let us examine the electromagnetic fields at a given space-time point
$x^\mu $. Using Maxwell equation
and the principle of causality, one can derive the retarded
Lienard-Weichert 4-potential (see [1], pp. 160-161; [2], pp. 654-656)
\begin{equation}
A_\mu = e\frac {v_\mu }{R^\alpha v_\alpha }.
\label{eq:LWPOTENTIAL}
\end{equation}
Here $v_\mu $ is the charge's 4-velocity at the retarded time and
$R^\mu $ is the 4-vector from the retarded space-time point to the
field point $x^\mu $. This 4-potential defines the fields uniquely.

A gauge transformation of $(\!\!~\ref{eq:LWPOTENTIAL})$
is (see [1], pp. 49-50; [2], pp. 220-223)
\begin{equation}
A'_\mu = A_\mu + \Phi {,_\mu} .
\label{eq:GAUGE}
\end{equation}
In the following lines, the form of the gauge function $\Phi (x^\mu)$
is investigated.

Relying on the variational principle, one finds constraints on
terms of the Lagrangian density. An examination of the Lagrangian
density $(\!\!~\ref{eq:LAGR})$, proves that every term of this
expression is a Lorentz scalar having the dimensions
$[L^{-4}]$. Thus, the action is a Lorentz
scalar and is dimensionless in the unit system
used here. In particular, the 4-potential $A_\mu $ must be a
4-vector whose dimension is $[L^{-1}]$. This requirement is
satisfied by the Lienard-Weichert 4-potential $(\!\!~\ref{eq:LWPOTENTIAL})$.
Thus, also $\Phi _{,\mu }$ of $(\!\!~\ref{eq:GAUGE})$
is a 4-vector and $\Phi $ must be a
dimensionless Lorentz scalar function of space-time coordinates.

Now, the coordinates are entries of a 4-vector. Therefore, the
general form of a homogeneous function which is a Lorentz
scalar depending on the coordinate must be a 
sum of power functions
of the form 
\begin{equation}
f_{a,p}(x^\mu ) = [(x^\mu - x_a^\mu)(x_\mu - x_{a\mu})]^p.
\label{eq:POWER}
\end{equation}
Here $2p$
denotes the order of the homogeneous function and $x_a^\mu $
denotes a specific space-time point. Relying on
the homogeneity of space-time, one finds that in the case discussed
here there is just one specific point $x_a^\mu$, which is the
retarded position of the charge. Thus, in order to be a
dimensionless Lorentz scalar, $\Phi $ must take the form
\begin{equation}
\Phi (x^\mu ) = \frac {c_1[(x^\mu - x_a^\mu)(x_\mu - x_{a\mu})]^p}
{c_2[(x^\mu - x_a^\mu)(x_\mu - x_{a\mu})]^p} = const.
\label{eq:PHI}
\end{equation}
Here the factors, $c_i$ denote numerical constants.

These arguments complete the proof showing that the gauge function $\Phi $
is a constant and the gauge 4-vector $\Phi _{,\mu }$ vanishes identically.
Hence, the Lienard-Weichert 4-vector $(\!\!~\ref{eq:LWPOTENTIAL})$ is
unique.

The foregoing result indicates the difference between an electrodynamic
theory where Maxwell equations and the Lorentz law of force are regarded
as the theory's cornerstone and a theory based on the variational
principle together with
causality and space-time homogeneity. Indeed, if Maxwell
equations are the theory's cornerstone then
it is very well known that one is free to
define the gauge function $\Phi(x^\mu )$ of $(\!\!~\ref{eq:GAUGE})$
(see [1], pp. 49-50; [2], pp. 220-223). For this reason, the result of this
work proves that the two approaches are {\em not equivalent.}


\newpage
References:
\begin{itemize}
\item[{*}] Email: elic@tauphy.tau.ac.il  \\
\hspace{0.5cm}
           Internet site: http://www-nuclear.tau.ac.il/$\sim $elic
\item[{[1]}] L. D. Landau and E. M. Lifshitz, {\em The Classical
Theory of Fields} (Pergamon, Oxford, 1975).
\item[{[2]}] J. D. Jackson, {\em Classical Electrodynamics} (John Wiley,
New York,1975).

\end{itemize}

\end{document}